\documentclass[12pt]{iopart}

\usepackage{iopams}  
\usepackage{graphicx}
\begin{document}

\title[Indirect Measurement Process Parameterized by ...]
{Indirect Measurement Process Parameterized by Nonunitary System - Pointer Interaction}

\author{Trifanov A.I., Miroshnichenko G.P.}

\address{Saint Petersburg National Research University of Information Technologies, 
Mechanics and Optics \\ 197101, Kronverksky 49, Saint Petersburg, Russia}
\ead{alextrifanov@gmail.com}
\begin{abstract}
An indirect quantum state measurement process is under investigation. Common evolution 
of detector's microscopic part (pointer) and measured system (target) is considered to 
be nonunitary due to the interaction between pointer and environment. System of 
differential equations for superoperators (operations), which map initial target state 
to it's conditional final state is obtained. Developed formalism is applied to the problem 
of cavity quantum mode photodetection to investigate basic information characteristics 
of this measurement process.
\end{abstract}

\maketitle

\section{Introduction}
Correct description of quantum measurements \cite{Braginsky} is one of the fundamental problems of 
Quantum Physics, which remains important and difficult research guidelines at present. 
Superselection and decoherentisation process, which closely related to this subject, control 
information flows in quantum communication systems \cite{Zurec}. There are at least three 
formulations, which reflect detector back action on the state of measured system \cite{CavesII}.

Indirect measurements use interaction between monitored (target) and ancillary (pointer) system  
followed by projection measurement on ancilla. This gives the wide opportunity in realizing indirect 
(and unsharp) detection process with weak disturbance of initial state of the target. Formally, this 
process may be described by parameterization $\left( {\rho_P ,U_{\tau},\left\{ {\Pi _x } \right\}}\right)$
of initial projective measurement. Here $\rho _P $ is initial pointer state, $U_{\tau}$ is evolution 
operator, $\tau$ is interaction time and $\left\{ {\Pi _x } \right\}$ are projector valued measures. 
It may be completed by canonical Naimark extension \cite{Naimark}. This technique were successfully 
applied in \cite{Sparaciari} to describe Stern-Gerlah experiment with unsharp measurement. Another way is 
to start from the arbitrary suitable pointer and comparatively simple interaction and construct 
Positive Operator Valued Measures (POVM) corresponding to this kind of generalized detection.
This method were used for description  of indirect measurement on trapped ions \cite{Choudhary} realizing 
quantum control and for quantum gate realization in QED cavity \cite{TrifanovI}. It should be noticed 
that unitary evolution is usually used for parameterization. 

In fact, the efficiency of measurement device is not exactly equal to identity (not to 
speak of the photodetection and particularly IR region, where this quantity is quite far away 
from unity). One can pronounce at least two factors, which participating in this process: 
the first one is a classical and quantum stochastic process, which govern the behavior of 
measurement apparatus. This factor may by taken into account by introducing the corresponding 
phenomenological probability distributions for detection events and errors \cite{TrifanovII}. 
The second one is influence of environment, which may appears in initially entangled state 
with the pointer and/or pointer state decoherentization during it's interaction with measured system.   

In this paper we investigate the conditional evolution of the monitored system in the case 
of interaction between pointer and environment during measurement process. Initial states of all 
subsystems are assumed to be factorized. In section 2 the brief review of required formalism 
is presented. Section 3 contains a simple example for developed theory and consider the 
case of evolution generated by superoperator in Linblad form. Namely, we deal with single 
cavity quantum mode photodetection with two level atom-pointer in use. In section 4 basic 
information characteristics, such as probability of certain state detection, information gain and 
fidelity for this measurement process are analyzed. Section 5 concludes the paper.  
  
\section{Description of parameterized measurement}

Interaction between pointer and environment leads to nonunitary common reduced evolution of 
target and pointer. To take it into account we will introduce the triplet 
$\left( {\rho _P ,\mathfrak{A}_{\tau},\left\{ {\Pi _x } \right\}} \right)$, 
where $\mathfrak{A}_{\tau} = \exp\left({- i \mathfrak{L} \tau}\right)$ and $\mathfrak{L}$ 
is evolution generator:
\begin{equation}
 i\dot \rho\left({t}\right)  = \mathfrak{L} \rho \left({t}\right).
\label{eq:2.1}
\end{equation}
Here $\rho\left({t}\right)$ is common density operator for system and pointer. Let 
$\rho_{in} = \rho\left({0}\right)$ and
\begin{equation}
\rho_{in} = \rho_S \otimes \rho_P = \rho_S \otimes \left| {\left. in \right\rangle } \right\rangle,
\label{eq:2.2}
\end{equation}
where $\left| {\left. in \right\rangle } \right\rangle =  \left|{in}\right\rangle \left\langle{in}\right|$ 
is initial pure pointer state in superoperator representation. Density operator 
$\mathfrak{A}_{\tau} \rho_{in} = \rho_{out}$ describe the common reduced state just 
after the interaction and before projective measurement.

If detection results are ignored the state of the monitored system may be obtained by tracing 
out ancillary degrees of freedom:
\begin{equation}
\rho'_S = Tr_A\left[{\rho_{out}}\right]
\label{eq:2.3}
\end{equation}
In the case of measurements with postselection an outcome space $\mathfrak{B}_A$ of pointer state 
detector is introduced. Each measurement result $r \in \mathfrak{B}_A$ corresponds to projector 
valued measure $\Pi_r$ on the pointer state space and superoperator $\Lambda_r$ which maps 
initial system state $\rho_{S}$ onto conditional final state $\rho_{S}^{r}$:
\begin{equation}
	\rho_{S}^{r} = \Lambda_r \rho_{S} / p_r,
\label{eq:2.4}
\end{equation}
where $p_r = \Tr_S \left({\Lambda_r \rho_{S}}\right)$ is probability of outcome $r$ and $\Tr_S$ is 
trace over state space of measured system. Alternatively, density operator $\rho_{S}^{r}$ may also 
be obtained from
\begin{equation}
	\rho_{S}^{r} = Tr_A\left[{\rho_{out} \Pi_r}\right].
\label{eq:2.5}
\end{equation}
In the following we will assume, that a secular approximation is established. Namely, 
typical time scale of intrinsic evolution of the pointer is large compared to time over 
which it's state varies appreciably. In this approach, if 
$\Pi_r = \left|{\varphi_r}\right\rangle \left\langle {\varphi_r}\right|$,
 $\Lambda_r$ may be written as
\begin{equation}
	\Lambda_r  =  \left\langle {\left\langle \varphi_r \right.} \right| \mathfrak{A}_\tau \left| 
	{\left. in \right\rangle } \right\rangle.
\label{eq:2.6}
\end{equation}

Secular approximation allows the following decomposition of superoperator $\mathfrak{A}_\tau$:
\begin{equation}
	\mathfrak{A}_\tau = \sum_{r,s \in \mathfrak{B}}{\left| {\left. \varphi_s \right\rangle } 
	\right\rangle \left\langle {\left\langle \varphi_r \right.} \right| \mathfrak{M}_{s,r} 
	\left({\tau}\right)}, 
\label{eq:2.7}
\end{equation}
where $\mathfrak{M}_{s,r} = \Lambda_r$ for initial conditions, indexed by $s \in \mathfrak{B}_A$. 
Substitution equation (\ref{eq:2.7}) into (\ref{eq:2.1}) gives the following system of 
differential equations for superoperators $\mathfrak{M}_{s,r}$:
\begin{equation}
i \dot{\mathfrak{M}}_{s,r} = \sum_{q \in \mathfrak{B}}{\left\langle {\left\langle \varphi_s \right.} \right| 
 \mathfrak{L} \left| {\left. \varphi_q \right\rangle } \right\rangle} \mathfrak{M}_{q,r}.
\label{eq:2.8}
\end{equation}
For every fixed $s$ we obtain the closed system for $\mathfrak{M}_{s,r}$. In the next section 
this formalism will be applied to the indirect photodetection problem. 

\section{Indirect photodetection process of cavity mode}

Indirect cavity mode photodetection is organized as follows. Two level atom passes through 
the cavity and interacts with excited quantum mode. Just after the interaction the atomic state 
is determined in a selective detector, which returns one of two possible alternatives: the atom 
is found in its ground state $\left| g \right\rangle _A $ or in it's exited state 
$\left| e \right\rangle _A $. Master equation in this case has Linblad form: 
\begin{equation}
i\dot \rho  = \mathfrak{L} \rho  = \left[ {H_{int} ,\rho } \right] + 
i \mathfrak{D}_A \rho. 
\label{eq:3.1}
\end{equation}
where $H_{int}$ is Jaynes-Cummings Hamiltonian:
\begin{equation}
H_{int}  = \Omega \sigma_{+} a \exp \left( {i\Delta t} \right) + \Omega ^* \sigma_{-} a^\dag  
\exp \left( { - i\Delta t} \right),
\label{eq:3.2}
\end{equation}
and $\mathfrak{D}_A$ is atomic "dissipater" :
\begin{equation}
\mathfrak{D}_A \rho  = \gamma _{ge}/2 \left( {2\sigma _{-}  \rho \sigma_{+} - \sigma_{+}  \sigma_{-} \rho - 
\rho \sigma_{+} \sigma_{-}  }\right) + \gamma _{eg}/2 \left( {c.c.} \right).
\label{eq:3.3}
\end{equation}
Here the following notations are used: $\Omega$ is a coupling between atom and mode, 
$\sigma_{+} = \left| e \right\rangle \left\langle g \right|$ and $\sigma_{-} = \left| g \right\rangle 
\left\langle e \right|$ are atomic operators, $a$ is an annihilation operator of cavity mode, 
$\Delta$ is a detuning, $\gamma_{ge}$ and $\gamma_{eg}$ are atomic population relaxation rates. 

Let us $\mathfrak{B}_A = \left\{{g,e}\right\}$ is a set of results of atomic state detection. Here we 
will assume the ideal (projective) measurement of pointer state. Using density operator decomposition 
\begin{equation}
\rho _{AF} \left( t \right) = \sum_{\mu, \nu \in \mathfrak{B}_A}{\left|{\mu}\right\rangle
\left\langle {\nu}\right|} \otimes \rho_{\mu\nu}\left({t}\right),
\label{eq:3.4}
\end{equation}
and secular approximation (rotating wave approximation in this case) $\Gamma \gg \Omega$, 
where $\Gamma \geq \left({\gamma_{eg}+\gamma_{ge}}\right)/2$ is atomic decoherentization rate, 
gives the following closed system of differential equations for 
$\rho_{gg} \left({t}\right)$ and $\rho_{ee} \left({t}\right)$:
\numparts
\begin{eqnarray}
\fl i \dot \rho _{gg}  = \kappa \Delta \left[ {a^\dag  a,\rho _{gg} } \right] + 
 i \kappa \Gamma \left( {2a^\dag  \rho _{ee} a - \rho _{gg} a^\dag  a - a^\dag  a\rho _{gg} } \right) + 
 i \left({\gamma _{eg} \rho _{ee}  - \gamma _{ge} \rho _{gg}}\right), 
 \label{eq:3.5a}\\
\fl i \dot \rho _{ee}  = \kappa \Delta \left[ {a^\dag  a,\rho _{ee} } \right] + 
 i \kappa \Gamma \left( {2a\rho _{ee} a^\dag   - \rho _{gg} aa^\dag   - aa^\dag  \rho _{gg} } \right) + 
 i \left({\gamma _{ge} \rho _{gg}  - \gamma _{eg} \rho _{ee}}\right),
\label{eq:3.5b}
\end{eqnarray}
\endnumparts
which may be written as:
\begin{equation}
 \dot{\rho}^d \left({t}\right) = \mathfrak{L}' \rho^d\left({t}\right), 
\label{eq:3.6}
\end{equation}
where $\rho^{d}\left({t}\right) = diag\left\{{\rho_{gg}\left({t}\right), 
\rho_{ee}\left({t}\right)}\right\}$ and $\kappa = {\left| \Omega  \right|^2 } / 
{\left( {\Gamma ^2  + \Delta ^2 } \right)}$. From equation (\ref{eq:2.8}) for $\mathfrak{L}'$ 
two separate systems of differential equations may be found. One of them correspond to atom prepared 
in ground state $\left|{g}\right\rangle_A$ and another to excited state $\left|{e}\right\rangle_A$.
For the first one:
\numparts
\begin{eqnarray}
 \dot{\mathfrak{M}}_{gg} = - \left( {\alpha K_0  + \beta  + \gamma _{ge} } \right) \mathfrak{M}_{gg} + 
 \left( {\alpha K_ +   + \gamma _{eg} } \right) \mathfrak{M}_{eg},
 \label{eq:3.7} \\
 \dot{\mathfrak{M}}_{eg} = \left( {\alpha K_ -   + \gamma _{ge} } \right) \mathfrak{M}_{gg} - 
 \left( {\alpha K_0  - \beta  + \gamma _{eg} } \right)\mathfrak{M}_{eg}.
\label{eq:3.8}
\end{eqnarray}
\endnumparts
Here $\alpha  = \kappa \Gamma$, $\beta  = \kappa \left( {i\Delta N - \Gamma/2} \right)$, 
$K_0 \rho_F  = \frac{1}{2}\left( {a^\dag  a\rho_F  + \rho_F aa^\dag  } \right)$,
$K_ +  \rho_F  = a^\dag  \rho_F a$, $K_ -  \rho_F  = a\rho_F a^\dag$, 
$N\rho_F  = \left[ {a^\dag  a,\rho_F } \right]$. Notice, $\left\{{K_0, K_+, K_-}\right\}$ 
form the full set of $SU\left( {1,1} \right)$ generators. System (\ref{eq:3.7} - \ref{eq:3.8}) 
may be solved analytically at least in two approximations: strong 
($\gamma _{ge}  < \kappa \Gamma  \approx \kappa \Delta  \ll \gamma _{eg} $) 
and weak  ($\gamma _{ge}  < \gamma _{eg}  \ll \alpha  \approx \kappa \Delta $) population 
relaxation limits. For the following we will use a numerical solution for this extreme cases.

\begin{figure}[t]
\centering\begin{tabular}{cc}
\includegraphics[width=17cm]{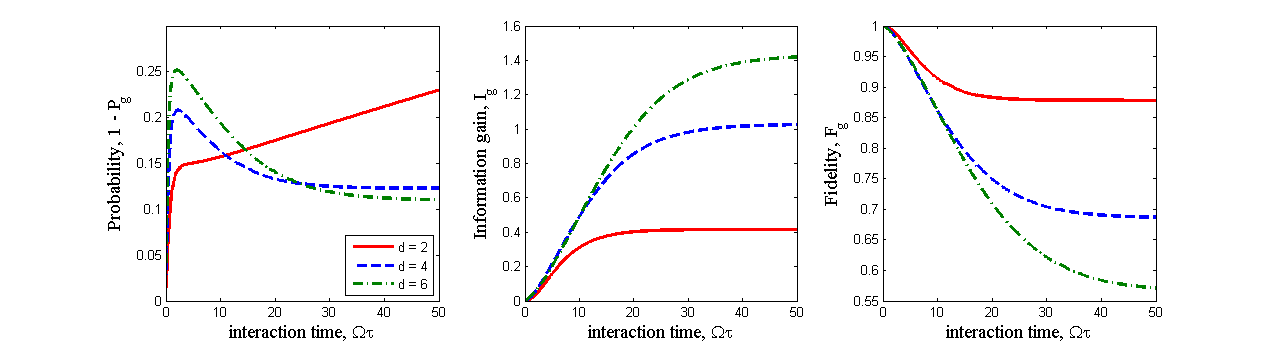}\\
\includegraphics[width=17cm]{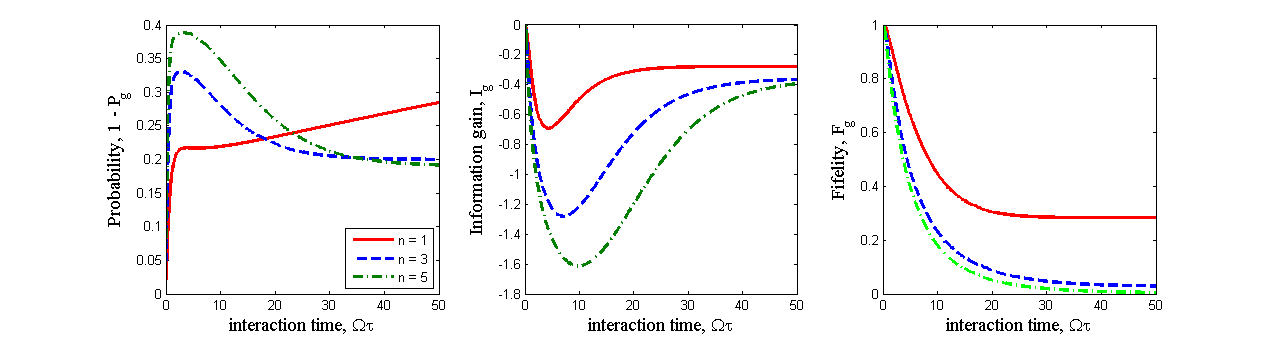}\\
\end{tabular}
\caption{Ground state detection: probability, information gain and fidelity as 
a functions of interaction time: for mixed state $\rho_F = 1/d$ (top) and 
Fock state $\rho_F = \left|{n}\right\rangle \left\langle {n}\right|$ (bottom). 
Strong relaxation approximation.}
\label{fig:1}
\end{figure}

\section{Information characteristics of photodetection process}

Here the basic information characteristics of the photodetection process will be 
investigated in bounds of the model discussed above. The following quantities 
are of most interest: \textit{probability} of ground or excited atomic state 
detection $P_r\left({t}\right) = Tr_F\left[{\mathfrak{M}_{gr}\left({t}\right) 
\rho_F}\right]$; \textit{information gain} $I_r = - \Delta H$ as a measure of 
entropy change $\Delta H = \rho_F^r \log \rho_F^r - \rho_F \log \rho_F$ 
resulted from photodetection; and \textit{fidelity} $F_r = \sqrt{\sqrt{\rho_{F}} 
\rho^{r}_{F} \sqrt{\rho_{F}}}$, which characterizes state change caused by 
measurement process.

\begin{figure}[t]
\centering\begin{tabular}{cc}
\includegraphics[width=17cm]{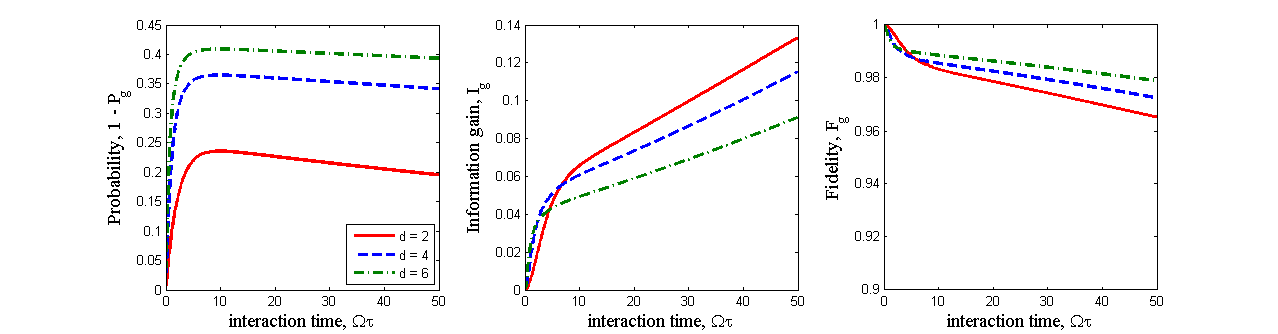}\\
\includegraphics[width=17cm]{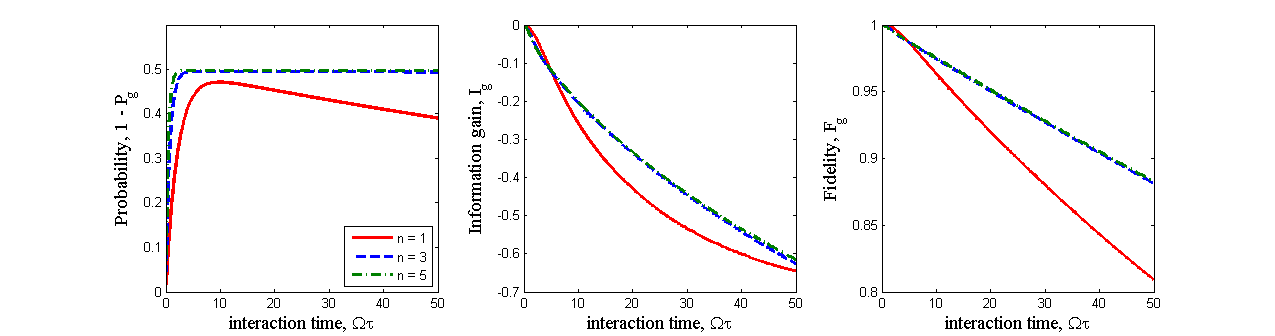}\\
\end{tabular}
\caption{Ground state detection: probability, information gain and fidelity as 
a functions of interaction time: for mixed state $\rho_F = 1/d$ (top) and 
Fock $\rho_F = \left|{n}\right\rangle \left\langle {n}\right|$ (bottom). 
Weak relaxation approximation.}
\label{fig:2}
\end{figure}

These quantities are shown on Fig.\ref{fig:1} and Fig.\ref{fig:2} as a functions 
of time interaction $\tau$ for different dimensions of cavity mode state space. Figure \ref{fig:1} 
shows the results obtained from strong relaxation approximation, while figure \ref{fig:2} 
shows these dependences for the case of weak relaxation. For each approximation two initial 
states are tested: completely mixed state $\rho_F = 1/d$ (top), and Fock state 
$\rho_F = \left|{n}\right\rangle \left\langle {n}\right|$ (bottom). Here 
$d$ is dimension of cavity mode state space and $n \leq d$ is a number of photons in 
mode (cases $d = 2, 4, 6$ and $n = 1, 3, 5$ are depicted).

For numerical modeling the following values of parameters is used (in unit $\gamma_{eg} = 1$): 
$\gamma_{ge} = 0.1$, $\gamma_{eg} = 1$, $\Gamma = 2$, $\Delta = 0.5$ and $\Omega = 0.7$ 
(for strong relaxation limit); $\gamma_{ge} = 0$, $\gamma_{eg} = 0.01$, $\Gamma = 2$, 
$\Delta = 0.5$ and $\Omega = 0.7$ (for weak relaxation limit). 

\section{Conclusion} 

The parameterization of measurement process in the presence of nonunitary evolution 
was investigated. Interaction between pointer and environment was discussed and applied 
to the quantum photodetection problem. Basic characteristics of measurement process as a 
functions of interaction time for two approximations were obtained. Properties of 
superoperators corresponded to different measurement result may be used for finding 
special regimes of detection: measurement without state or entropy change from one side 
and detection with best information gain from another. The generalization of previous 
description beyond secular approximation (quantum Brownian motion) will be considered in 
future. It is attractive to obtain detector superoperators from evolution generator, 
though it requires deep analyze of their algebraic properties.

\end{document}